\documentclass[twocolumn,preprintnumbers,amsmath,amssymb,prl,floatfix]{revtex4}

\usepackage{graphicx,color}
\usepackage{dcolumn}
\usepackage{bmpsize}
\usepackage{bm}

\begin{document}

\title{Quantum and Thermal Phase Transitions of the Triangular SU(3) Heisenberg Model under Magnetic Fields}

\author{Daisuke Yamamoto}
\email{d-yamamoto@phys.aoyama.ac.jp}
\author{Chihiro Suzuki}
\author{Giacomo Marmorini}
\author{Sho Okazaki}
\author{Nobuo Furukawa}
\affiliation{Department of Physics and Mathematics, Aoyama Gakuin University, Sagamihara, Kanagawa 252-5258, Japan}
\date{\today}
\begin{abstract}
We study the quantum and thermal phase transition phenomena of the SU(3) Heisenberg model on triangular lattice in the presence of magnetic fields. Performing a scaling analysis on large-size cluster mean-field calculations {endowed} with a density-matrix-renormalization-group solver, we reveal the quantum phases selected by quantum fluctuations from the massively degenerate {classical} ground-state manifold. The magnetization process up to saturation reflects three different magnetic phases.   {The low- and high-field phases have strong nematic nature, and especially the latter is found only via a nontrivial reconstruction of symmetry generators from the standard spin and quadrupolar description.} We also perform a semi-classical Monte-Carlo simulations to show that thermal fluctuations prefer the same three phases as well. Moreover, we find that exotic topological phase transitions driven by the binding-unbinding of fractional (half-)vortices take place, 
  {due to the nematicity of the low- and high-field phases}. Possible experimental realization with alkaline-earth-like cold atoms is also discussed. 
\end{abstract}
\maketitle

{\it Introduction.}--- 
In solid-state physics, lattice Hamiltonians symmetric under the special unitary group of degree $\mathcal{N}=2$, denoted by SU(2), have been intensively studied since the electron -- the main actor in solids -- has two internal (spin) degrees of freedom. Higher degree of symmetry, or $\mathcal{N}>2$, can be accessed only with fine-tuning of parameters in some models, e.g., of spin liquid crystals~\cite{nakatsuji-05,tsunetsugu-06,bhattacharjee-06} and transition metal oxides~\cite{kugel-73,arovas-95,li-98,tokura-00}, or as a consequence of exotic emergent phases~\cite{keller-14,chen-15,quito-20}. However, recent advances in experiments with cold gases of alkaline-earth(-like) atoms, such as $^{173}$Yb~\cite{fukuhara-07,cazalilla-09,hara-11,taie-12,mancini-15,hofrichter-16,ozawa-18} and $^{87}$Sr~\cite{desalvo-10,tey-10}, have provided a new platform and strong motivation in studying the enhanced continuous symmetry of SU($\mathcal{N}>2$). Since those atoms possess symmetric interactions {under} nuclear spin $\bm{I}$  ($I=5/2$ for Yb and $9/2$ for Sr), loading them into optical lattices   {enables us to} create an ideal quantum simulator of the {SU($\mathcal{N}\leq 2I+1$)} extension of the Hubbard model~\cite{honerkamp-04} and its strong-coupling limit, namely the SU($\mathcal{N}$) Heisenberg model~\cite{gorshkov-10,nataf-14}. In such higher symmetric systems, the ground states often form a massively (quasi)degenerate manifold. Therefore, of particular interest are the quantum and thermal fluctuations selecting one of the many-body states and the emergence of exotic phase transition phenomena~\cite{lacroix-11}.

The SU(3) Heisenberg model on triangular lattice has been theoretically studied as a special symmetric point of the spin-1 {bilinear-biquadratic} model~\cite{lauchli-06,smerald-13,bauer-12}. 
Since the number of colors ($\mathcal{N}=3$) is compatible with the tripartite structure of the triangular lattice, the SU(3) Heisenberg model with antiferromagnetic couplings exhibits no (apparent) geometrical frustration, unlike the SU(2) case~\cite{moessner-06}.   {Indeed, the ground state is uniquely determined (up to trivial degeneracy) to be a simple three-color three-sublattice order at the level of the classical, {mean-field}, analysis~\cite{lauchli-06,smerald-13}, and it has been confirmed by numerical investigations~\cite{bauer-12}. 
Whereas the ground state may not be so exciting, the properties under the presence of magnetic field remain an interesting open problem since the {mean-field} analysis yields an accidental continuous degeneracy~\cite{lauchli-06}. 

  {In this Letter, we explore the effect of quantum and thermal fluctuations on the phase transition phenomena of the triangular SU(3) Heisenberg model in magnetic fields.} High magnetic field experiments have been playing a central role in understanding the properties of magnetic materials~\cite{berthier-02}, one of the fundamental reasons being that a magnetic field, in combination with  lattice geometry, topological features, fluctuation effects, etc., stimulates the emergence of a rich variety of nontrivial magnetic states such as magnetization plateaus~\cite{chubukov-91,shirata-12,nishimoto-13}, nematic states~\cite{nawa-13,buttgen-14}, and field-induced quantum spin liquids~\cite{nishimoto-13,baek-17}. 
This is naturally expected to occur for general SU($\mathcal{N}$) systems. 

  {First}, we employ the cluster mean-field plus scaling (CMF+S) method~\cite{yamamoto-12-2,yamamoto-14,yamamoto-17} with two-dimensional (2D) density matrix renormalization group (DMRG) solver~\cite{yamamoto-19} to reveal the quantum phases selected from the nontrivial classical ground-state manifold. We find that the quantum order-by-disorder mechanism stabilizes three different phases 
depending on the field strength, until the system reaches the magnetic saturation. 
{Of particular significance is that,} although the high-field (HF) phase appears to be a conventional (nonnematic) spin order in terms of the spin and quadrupolar operators, we reveal a concealed nematic nature by 
reconstructing the symmetry generators. 
  {Furthermore}, we develop a framework of {{\it semiclassical} multicolor Monte Carlo} simulations~\cite{stoudenmire-09} by introducing a ``relaxation acceleration'' technique, and discuss the thermal phase transition phenomena. In addition to the stabilization of the same three phases by thermal fluctuations, we find particular topological phase transitions characterized by the binding-unbinding of fractional (half-)vortices.

{\it The SU(3) Heisenberg model in magnetic fields.}--- 
The SU(3) Heisenberg model is given by
\begin{eqnarray}
\hat{\mathcal{H}}_{\rm SU(3)}=2J\sum_{\langle i,j\rangle}\sum_{{\rm A}=1,2,\cdots,8}\hat{T}_i^{\rm A} \hat{T}_j^{\rm A}\label{hamiltonian}~~(J>0),
\end{eqnarray}
where $\hat{T}_i^{\rm A}=\hat{\lambda}_i^{\rm A}/2$ are the eight generators of the SU(3)  Lie algebra  in the defining representation. To draw connections to the spin physics, here we employ the spin-1 operator $\hat{\bm{S}}_i=(\hat{S}^x_i,\hat{S}^y_i,\hat{S}^z_i)$   {for ${\rm A}=1,2,3$} and the quadrupolar operator $\hat{\bm{Q}}_i=(\hat{Q}^{x^2-y^2}_i,\hat{Q}^{z^2}_i,\hat{Q}^{xy}_i,\hat{Q}^{yz}_i,\hat{Q}^{xz}_i)$   {for ${\rm A}=4,\cdots,8$} as $\hat{\lambda}_i^{{\rm A}}$, instead of the    {standard} Gell-Mann matrix {basis}. The quadrupolar operators are  $(\hat{S}^x_i)^2-(\hat{S}^y_i)^2$, $\sqrt{3}(\hat{S}^z_i)^2-2/\sqrt{3}$, $\hat{S}^x_i\hat{S}^y_i+\hat{S}^y_i\hat{S}^x_i$, $\hat{S}^y_i\hat{S}^z_i+\hat{S}^z_i\hat{S}^y_i$, and $\hat{S}^z_i\hat{S}^x_i+\hat{S}^x_i\hat{S}^z_i$, respectively. In this spin-1 representation, the Hamiltonian~(\ref{hamiltonian}) is equivalent to {the {bilinear-biquadratic} model~\cite{lacroix-11,lauchli-06,smerald-13,bauer-12,toth-12,niesen-18} with equal {positive} coefficients,   {acting}  on spin states $\sigma=-1,0,1$:}
\begin{eqnarray}
\hat{\mathcal{H}}_{\rm SU(3)}=\frac{J}{2}\sum_{\langle i,j\rangle}\left(\hat{\bm{S}}_i\cdot\hat{\bm{S}}_j+\hat{\bm{Q}}_i\cdot\hat{\bm{Q}}_j\right). \label{hamiltonianS}
\end{eqnarray}

Below, we discuss the system under magnetic (Zeeman) fields: $\hat{\mathcal{H}}\equiv \hat{\mathcal{H}}_{\rm SU(3)}+\hat{\mathcal{H}}_{\rm Z}$ with $\hat{\mathcal{H}}_{\rm Z}=-H\sum_{i}\hat{S}_i^z$. The magnetic field explicitly breaks the SU(3) symmetry down to U(1)$\times$U(1); {specifically}, the global rotations around the $\hat{S}^z$ and $\hat{Q}^{z^2}$ axes [hereafter, written as U(1)$_{S^z}$ and U(1)$_{Q^{z^2}}$] remain since $\sum_i [\hat{S}_i^z,\hat{\mathcal{H}}]=\sum_i [\hat{Q}_i^{z^2},\hat{\mathcal{H}}]=0$. 
{Within the site-decoupling {mean-field} approximation~\cite{lauchli-06}, the specific spin and quadratic orders in the ground state for $0<H<H_{\rm s}$ (with $H_{\rm s}=9J$) exhibit a massive,   {accidental} degeneracy not related to the symmetries of the system.}   {The detailed structure of the degenerate ground-state manifold is described in the Supplemental Material~\cite{SM}.}

\begin{figure}[t]
\includegraphics[scale=0.47]{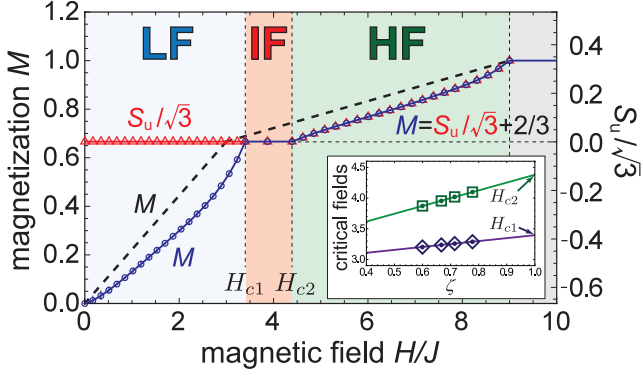}
\caption{\label{fig1}
Field dependences of the magnetization $M$ (blue circles) and {the uniform scalar nematic order parameter $S_{\rm u}$} (divided by $\sqrt{3}$, red triangles), obtained by the CMF+S analysis. The {classical (mean-field)} value of $M$ is plotted together (dashed line). Left and right axes are shifted by 2/3. The inset shows cluster-size scalings of the critical fields. 
}
\end{figure}
{\it Quantum order by disorder.}--- 
In order to discuss the lifting of the accidental degeneracy by quantum fluctuations, we perform the CMF+S calculations~\cite{yamamoto-12-2,yamamoto-14,yamamoto-17} with 2D DMRG solver~\cite{yamamoto-19}. We employ a triangular-shaped cluster of $N_{\rm C}$ sites, in which the quantum intersite correlations are treated {exactly} within the cluster, whereas the couplings with the outside spins are replaced by {mean-field} interactions.   {Under the three-sublattice ($\mu={\rm A},{\rm B},{\rm C}$) ansatz,}   {the self-consistent equations 
  {$
\langle\hat{\bm{S}}_\mu \rangle=\frac{3}{N_{\rm C}}\sum_{i_\mu \in {\rm C}}\langle\Psi_{{N_{\rm C}}}|\hat{\bm{S}}_{i_\mu} |\Psi_{{N_{\rm C}}} \rangle
$} 
and the {analogous} expressions for $\langle\hat{\bm{Q}}_\mu \rangle$ are solved by calculating the ground state of the $N_{\rm C}$-site cluster, $|\Psi_{{N_{\rm C}}} \rangle$, with 2D DMRG in an iterative way until convergence~\cite{yamamoto-19}.} The scaling parameter {$\zeta\equiv N_{\rm B}/(3N_{\rm C})$}, with $N_{\rm B}$ being the number of bonds inside the cluster, serves as an indicator of the extent to which quantum correlations are taken into account, interpolating the classical ({$N_{\rm C}=1$}; $\zeta=0$) and exactly-quantum ($N_{\rm C}\rightarrow \infty$; $\zeta=1$) limits. Here we perform the calcuations for $N_{\rm C}=10,15,21$ ($\zeta=3/5,2/3,5/7$) and make the {linear} extrapolation {of the results toward $\zeta\rightarrow 1$} with an error bar estimated from the derivation of different sets of cluster sizes used for the extrapolation. {The larger size cluster of $N_{\rm C}=36$ ($\zeta=7/9$) is also considered for the determination of the phase boundaries (see the inset of Fig.~\ref{fig1}).}

We plot the quantum magnetization curves {$M(H)\equiv \sum_\mu \langle \hat{S}^z_\mu \rangle/3$} obtained by the CMF+S in Fig.~\ref{fig1}. 
The low-field (LF) phase is characterized by $\langle \hat{S}^{z}_A\rangle=\langle \hat{S}^{z}_B\rangle\neq \langle \hat{S}^{z}_C\rangle\approx 0$, $\langle \hat{Q}^{x^2-y^2}_A\rangle=-\langle \hat{Q}^{x^2-y^2}_B\rangle$, {and} $\langle \hat{Q}^{x^2-y^2}_C\rangle=0$, modulo a global rotation in the ($Q^{x^2-y^2}$, $Q^{xy}$) plane and sublattice exchanges; the other components are all zero [see Fig.~\ref{fig2}(a)]. Although the spin sector $(S^x,S^y,S^z)$ forms a collinear structure along the field axis, the transverse quadrupolar moments $(Q^{x^2-y^2}$, $Q^{xy})$ break the rotational symmetry around $S^z$. {It is particularly interesting  that a $\pi$ rotation around the $S^z$ axis is sufficient for $(Q^{x^2-y^2},Q^{xy})$ to return the initial state} {as illustrated in Fig.~\ref{fig2}(b)} due to the nematic nature{, reflecting the factor 2 in the commutation relation $[\hat{Q}^{x^2-y^2},\hat{Q}^{xy}]=2i\hat{S}^z$}. Thus, it is concluded that the LF phase breaks the [U(1)$_{S^z}/\mathbb{Z}_2$]$\times \mathbb{Z}_3$ (i.e.,  half of the original rotational and threefold translational) symmetries. 
  {Consequently}, the remaining U(1)$_{Q^{z^2}}$ symmetry guarantees the preservation of the {uniform nematic scalar order parameter $S_{\rm u}\equiv \sum_\mu \langle \hat{Q}^{z^2}_\mu \rangle/3$}, resulting in the plateau formation   {at zero value} in Fig.~\ref{fig1}. 

\begin{figure}[t]
\includegraphics[scale=0.5]{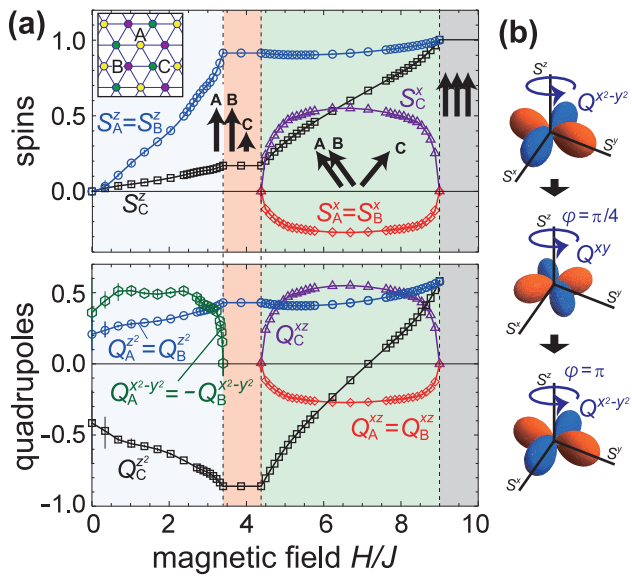}
\caption{\label{fig2}
(a) Nonzero components of the spin and quadrupolar moments, obtained by the CMF+S analysis in a fixed gauge with $\langle\hat{Q}_A^{xy} \rangle=\langle\hat{Q}_A^{yz} \rangle=0$. The inset shows the three-sublattice structure. (b) Spherical plots of $|\langle \bm{S}|\hat{Q}^{x^2-y^2} |\bm{S} \rangle|$ and its $\pi/2$ and $\pi$ rotations about $U(1)_{S^z}$ with $|\bm{S} \rangle$ being the spin coherent state pointing in the $\bm{S}$ direction~\cite{lacroix-11}.
}
\end{figure}
At $H=H_{c1}=3.40J$, the transverse quadrupolar moments vanish and the U(1)$_{S^z}$ symmetry is restored. Thus, in the intermediate-field (IF) phase, both $M$ and $S_{\rm u}$ exhibit plateau behavior in the range of $H_{c1}<H<H_{c2}=4.38J$. The longitudinal spin moments $\langle \hat{S}^{z}_\mu \rangle$ have the values of   {approximately $(1,1,0)$ (not exactly, due to  quantum depletion)} and thus $M= 2/3$. {Such a plateau formation has been reported in the spin-1 {bilinear-biquadratic} model {when} the quadrapolar coupling is larger than the dipolar one~\cite{lauchli-06}. Our results showed that the plateau is stabilized by purely quantum effects even for equal bilinear-biquadratic [SU(3)-symmetric] coupling.}

In the HF phase, the spin $(S^x,S^y,S^z)$ sector forms a ``2:1'' structure of the V shape, similar to the SU$(2)$ case~\cite{chubukov-91}.   {Therefore, it apparently seems to be a standard non-nematic spin order. However, we notice that the curves of $M$ and $S_{\rm u}/\sqrt{3}$ differ only by a constant shift of 2/3. We {show} that this feature stems} from a particular spontaneous  partial breaking of U(1)$_{S^z}\times$U(1)$_{Q^{z^2}}$: the linear combination of generators $\hat{P}^z_+\equiv \frac{1}{2}\hat{S}^z+\frac{\sqrt{3}}{2}\hat{Q}^{z^{2}}$ is broken, while $\hat{P}^z_-\equiv \frac{\sqrt{3}}{2}\hat{S}^z-\frac{1}{2}\hat{Q}^{z^{2}}$ is preserved. The U(1)$_{P^z_\pm}$ action produces a rotation of {the system} in the plane of $\hat{P}^x_\pm\equiv (\hat{S}^x\pm\hat{Q}^{xz})/\sqrt{2}$ and $\hat{P}^y_\pm \equiv (\hat{S}^y\pm\hat{Q}^{yz})/\sqrt{2}$.   {As is seen in Fig.~2(a), the transverse spin and quadrupolar moments hold the relation}   {$\langle \hat{P}^x_{-,\mu} \rangle=\langle \hat{P}^y_{-,\mu} \rangle=0$}   {in the HF phase, which indicates} the preservation of the U(1)$_{P^z_-}$ symmetry. As for   {the broken} U(1)$_{P^z_+}$, a $\pi$ rotation is sufficient for $(\hat{P}^x_+,\hat{P}^y_+)$ to return to the initial state since $[\hat{P}^x_+,\hat{P}^y_+]=2i{\hat{P}^z_+}$, and thus   {the HF phase possesses a nematic nature despite the apparent spin (dipolar) order.} Considering also the sublattice exchange, we conclude that the HF phase breaks [U(1)$_{P^z_+}/\mathbb{Z}_2$]$\times \mathbb{Z}_3$.

{The above results extend the widely believed conjecture~\cite{lacroix-11}, originally formulated for the standard SU(2) case, that the order-by-disorder selection mostly favors a ``collinear'' state with only diagonal components,  followed by ``coplanar'' states with the moment vectors 
on all sublattices lying in one plane that includes the rotation axis, since their fluctuations are softer.
Here we have demonstrated that this is true also in a model with underlying SU(3) symmetry (see also Ref.~\cite{SM} for the linear flavor-wave excitation spectra): the IF phase, having only diagonal order, is collinear, whereas the other two phases can be seen to be coplanar once the appropriate plane, containing the rotation axis (broken symmetry generator), in the SU(3) space is identified [the ($Q^{x^2-y^2}$,$S^z$) plane for LF and the ($P_+^x$,$P_+^z$) plane for HF in the gauge of Fig.~\ref{fig2}(a)].}

{\it Thermal phase diagram.}--- 
Given the strong nematic nature of the zero-temperature phases, it is interesting to study the thermal phase transitions, especially associated with the [U(1)/$\mathbb{Z}_2$]$\times \mathbb{Z}_3$ symmetry breaking. We employ the {semiclassical Monte Carlo} simulations~\cite{stoudenmire-09} within the direct-product {approximation: $ |\Psi^{\rm cl}\rangle =\otimes_i |\psi_i\rangle$ with local wave functions  $|\psi_i\rangle=\sum_{\sigma} d_{i,\sigma}|\sigma_i \rangle$ ($|\bm{d}_i|^2=1$).} The standard Metropolis updates are performed for the coefficients $d_{i,\sigma}$ on $L\times L$ rhombic clusters under periodic boundary conditions, based on the Boltzmann {distribution} $p\propto \exp (- E_{\rm cl}/k_{\rm B} T)$   {with $E_{\rm cl}\equiv \langle \Psi^{\rm cl}|\hat{\mathcal H}|\Psi^{\rm cl}\rangle$~\cite{stoudenmire-09}}. We further develop the {method} by applying a {``relaxation acceleration''} with local unitary transformations $e^{i c \hat{\mathcal{H}}^{\rm loc}_i}|\psi_i\rangle$, where $c$ are {uniformly distributed} random numbers and $\hat{\mathcal{H}}^{\rm loc}_i\equiv (\otimes_{j\neq i}\langle \psi_j|)\hat{\mathcal{H}}(\otimes_{j\neq i}|\psi_j\rangle)$. Here we choose, after some trials, $|c| \leq \pi\|\hat{\mathcal{H}}^{\rm loc}_i\|_{\rm F}^{-1}$ with $\|\cdots\|_{\rm F}$ being the Frobenius norm. The {relaxation-acceleration} sweeps over lattice sites are performed twice following each Metropolis update sweep. This method, applied to highly symmetric systems, is significantly more efficient in improving  decorrelation and avoiding trapping in local minima~\cite{SM}.

Figure~\ref{fig3} shows the thermal phase diagram obtained by the {semiclassical Monte Carlo method}, which is reliable in the region away from the low-temperature quantum regime, since it neglects the intersite quantum correlations.  It is seen that the same three (LF, IF, and HF) phases are selected also by thermal fluctuations from the classical degenerate manifolds at $T=0$. 
The boundaries are determined by the divergence of the correlation length and the scaling analyses of the susceptibility for the corresponding components~\cite{SM}. 

\begin{figure}[t]
\includegraphics[scale=0.5]{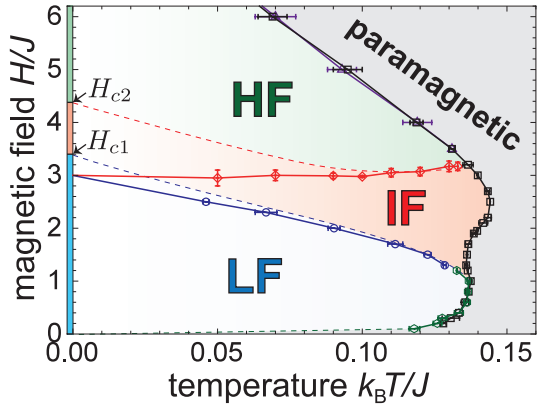}
\caption{\label{fig3}
Thermal phase diagram obtained by the semiclassical Monte Carlo simulations. We also mark the critical fields $H_{{\rm c}1}$ and $H_{{\rm c}2}$ obtained by the CMF+S method at the quantum ($T=0$) limit. The dashed lines are the sketches of the  phase boundaries expected from the combination of the semiclassical Monte Carlo (valid at high temperatures) and CMF+S (valid at $T=0$) results.
}
\end{figure}
We show in Fig.~\ref{fig4}(a) the stiffness $\rho_{S^z}(T)$ for a twist of the spin and quadrupolar moments around $S^z$ near the LF-IF transition. It is seen that $\rho_{S^z}(T)$ at the transition point $T=T_{\rm c}$ does not satisfy the standard universal relation $\rho_{S^z}(T_{\rm c})=2k_{\rm B}T_{\rm c}/\pi$ for the {Berezinskii-}Kosterlitz-Thouless transitions~\cite{kosterlitz-16}. This is attributed to the nematic nature of $(Q^{x^2-y^2},Q^{xy})$, which break  U(1)$/\mathbb{Z}_2$ rotations around $S^z$ [shown in Fig.~\ref{fig2}(b)]. Because of this, the $(Q^{x^2-y^2},Q^{xy})$ moments can form a topologically stable vortex with fractional vorticity $\rho_v=1/2$ [Fig.~\ref{fig4}(b)], unlike in the standard XY universality class, where $\rho_v=1$.   {This half-vortex is analogous to the 180$^\circ$ disclination of nematic liquid crystals~\cite{mermin-79}}.
The transition from LF to IF is associated with the unbinding of pairs of half-vortex and half-antivortex, resulting in the modified universal relation $\rho_{S^z}(T_{\rm c})=2k_{\rm B}T_{\rm c}/\pi \rho_v^2=8k_{\rm B}T_{\rm c}/\pi$~\cite{korshunov-02}{, which has been discussed also in spin-1 superfluids~\cite{mukerjee-06}}. 
This particular topological transition takes place also at the boundary of the HF and IF (or paramagnetic) phase [Fig.~\ref{fig4}(c)], where it is related to the U(1)$/\mathbb{Z}_2$ rotation around ${P^z_+}$ mentioned above. This universal jump is associated with the unbinding of half-vortex pairs in the ($P^x_+$,$P^y_+$) plane. 
\begin{figure}[t]
\includegraphics[scale=0.5]{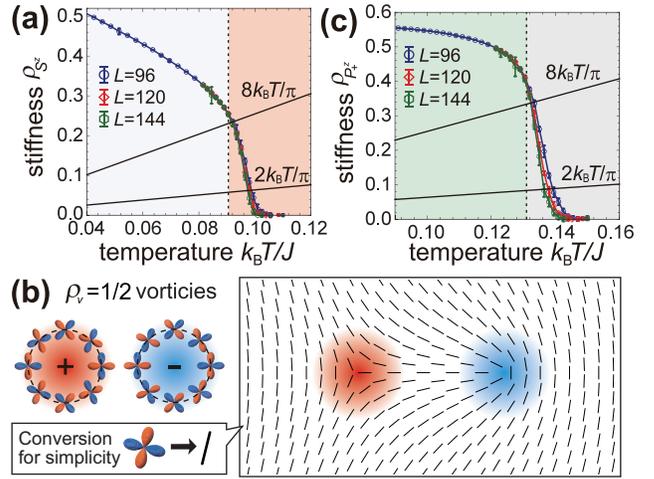}
\caption{\label{fig4}
(a) Stiffness $\rho_{S^z}(T)$ along $H/J=2.0$, which shows a universal jump $\rho_{S^z}(T_{\rm c})=8k_{\rm B}T_{\rm c}/\pi$ at the LF-IF transition,   {except for a slight finite-size effect}. (b) Vortex and antivortex with half-vorticity $\rho_v=\pm 1/2$ in the   {projected $(Q^{x^2-y^2},Q^{xy})$ plane. The right-hand panel is a schematic illustration of a topological half-vortex pair excitation on the background of a {\it uniform} quadrupolar order on, say, sublattice A.} (c) Same as in (a) for $\rho_{P^z_+}(T)$ at the HF-paramagnetic transition along $H/J=3.5$. }
\end{figure}

Let us comment briefly on the limit of $H=0$. 
Since the classical ground state is given by   {$\bm{d}_i=(1,0,0)$, $(0,1,0)$, and $(0,0,1)$ for sublattice A, B, and C, respectively, or SU(3) rotations thereof~\cite{smerald-13}}, the symmetry is spontaneously broken down to U(1)$\times$U(1). The fundamental group $\pi_1$[SU(3)/U(1)$\times$U(1)] is trivial~\cite{ueda-16} and therefore there are no vortex-induced finite temperature phase transitions~\cite{kosterlitz-16,kawamura-84}. The tendency of the IF-paramagnetic line toward $(T,H)=(0,0)$ corroborates this scenario.

{\it Experimental realization.}--- 
{A promising way for realizing the present system is picking up three nuclear spin components of alkaline-earth(-like) atoms, e.g., $I_z=-5/2,-1/2,3/2$ of $^{173}$Yb~\cite{fukuhara-07,cazalilla-09,hara-11,taie-12,mancini-15,hofrichter-16,ozawa-18}, as $\sigma=-1,0,1$ via the optical pumping. Without introducing overall imbalance in spin population, one could study the magnetic-field ($H$) effects by applying a state-dependent potential gradient, say, in the $x$ direction, $V^{\rm ext}_{\sigma}(x)=\sigma V x$, which realizes the magnetization process in $-H_s<H<H_s$ in real space as a function of the local magnetic field $H(x)=2Vx$ (in the sense of the local density approximation~\cite{bergkvist-04}). Such a potential gradient could be prepared by the combination of circularly and linearly polarized lights~\cite{ozawa-18} with 
a fine-tuning to keep the condition $\mu_1-\mu_0=\mu_{0}-\mu_{-1}=H$ for the local chemical potentials of each component. If, alternatively, one uses a real magnetic-field gradient, closed-shell alkaline-earth(-like) atoms do not suffer from quadratic Zeeman effects~\cite{jenkins-39} in the present field range $H\sim J$ and no fine-tuning is needed. Another, perhaps more efficient, way 
is the introduction of a coherent laser coupling between different spin states~\cite{mancini-15,fallani-20},
 since it can create a field term $-H\sum_i\hat{S}_i^x$ instead of $\hat{\mathcal{H}}_Z$, but all the results presented here remain valid up to a global spin  rotation.}

{The estimated critical temperature, $T/J\approx 0.14/k_{\rm B}$ at its highest value, is a realistic goal for the first observation of the SU($\mathcal{N}$) order by disorder, given that $T/J\approx 0.9/k_{\rm B}$ has been achieved in SU(2) systems~\cite{mazurenko-17}, considering also the Pomeranchuk cooling effect~\cite{taie-12} for many-component systems and the fact that the specific spin correlations can be detected from temperatures higher (typically 2-3 times~\cite{brown-17})  than the true critical temperature shown in Fig.~\ref{fig3}. The formation of the three-sublattice orders can be observed by the time-of-flight image of the momentum distribution~\cite{brown-17,parsons-16,boll-16,cheuk-16,mazurenko-17,hilker-17}, and the IF state would appear as a spatial plateau in the case of varying potential $H(x)$. The singlet-triplet oscillation~\cite{ozawa-18,trotzky-10} should exhibit different characteristic behaviors for each phase. In addition, the extension of the quantum-gas microscope technique to fermionic SU($\mathcal{N}$) systems~\cite{miranda-15,yamamoto-16} could provide {a wealth of detailed} measurements, including the formation of half-vortices. }

{A global spin population imbalance~\cite{brown-17} indirectly creates chemical potential differences among the components and, in general, an extra term $A\sum_i \hat{Q}_i^{z^2}$ has to be considered in addition to $H$. Exploring the entire $(H,A,T)$ space would be an interesting future subject. }

{\it Conclusions.}---
We studied the quantum and thermal phase transition phenomena of the SU(3) Heisenberg model under magnetic fields by using the CMF+S and semiclassical Monte Carlo methods. We demonstrated that pure quantum-fluctuation effects stabilize a magnetization plateau at 2/3 of the saturation in the intermediate range of the field strength. The uniform scalar nematic order parameter also forms a plateau at zero value, which, more interestingly, appears already in the lower-field phase with no magnetization plateau. 
The high-field phase exhibits {an unexpected nematic nature stemming from nontrivial partial breaking of U(1)$\times $U(1) symmetry.} Moreover, the strong nematic nature of the low- and high-field phases gives rise to fractional vortices and antivortices, whose pair dissociation results in a topological phase transition with vorticity $\rho_v=1/2$ at the critical temperature.

The above results, together with the calculated critical temperatures, provide a robust guideline for future experiments with alkaline-earth(-like) atoms. 
Additionally, the physics we explored is relevant to solid-state materials with nearly SU(3) symmetric parameters and, more generally, to systems with multipolar orders. {
In solids, a sizable spin-lattice coupling can in principle lock the quadrupolar orders to certain directions and lead to clock-type transitions at low temperatures; this kind of  phenomenon is clearly absent in the cold-atom setting.}


\begin{acknowledgments}
We thank Y. Takahashi and I. Danshita for valuable discussions on this subject. This work was supported by KAKENHI from Japan Society for the Promotion of Science, Grant No. 18K03525 (D.Y.), CREST from Japan Science and Technology Agency No.~JPMJCR1673 (D.Y.), and ``Early Eagle'' grant program from Aoyama Gakuin University Research Institute. 
\end{acknowledgments}

\onecolumngrid

\newpage 

\subsection{Supplementary Material for ``Quantum and Thermal Phase Transitions of the Triangular SU(3) Heisenberg Model under Magnetic Fields''}
\renewcommand{\thesection}{\Alph{section}}
\renewcommand{\thefigure}{S\arabic{figure}}
\renewcommand{\thetable}{S\Roman{table}}
\setcounter{figure}{0}
\newcommand*{\citenamefont}[1]{#1}
\newcommand*{\bibnamefont}[1]{#1}
\newcommand*{\bibfnamefont}[1]{#1}

\renewcommand{\theequation}{S\arabic{equation}}
\renewcommand{\thesection}{\Alph{section}}
\renewcommand{\thefigure}{S\arabic{figure}}
\renewcommand{\thetable}{S\Roman{table}}
\setcounter{equation}{0}

\subsection{\label{0}Classical degeneracy manifold of the triangular SU(3) Heisenberg model with magnetic fields}
Within the site-decoupling {mean-field} approximation, the ground state is assumed to be a direct product of local wave-functions ($|\bm{d}_i|^2=1$): 
\begin{eqnarray}
|\Psi^{\rm cl}\rangle =\otimes_i |\psi_i\rangle~~{\rm with}~~|\psi_i\rangle=\sum_{\sigma} d_{i,\sigma}|\sigma_i \rangle.\label{direct_product}
\end{eqnarray}
The coefficient vector $\bm{d}_{i}=(d_{i,-1},d_{i,0},d_{i,1})$ normalized to unit length ($|\bm{d}_i|=1$) identifies the local state at site $i$ as a superposition of the three basis states ($\sigma=-1,0,1$). 
Under the three-sublattice ($\mu={\rm A},{\rm B},{\rm C}$) ansatz, the variational energy $E_{\rm cl}\equiv \langle \Psi^{\rm cl}|\hat{\mathcal H}|\Psi^{\rm cl}\rangle$ can be written as 
\begin{eqnarray}
\frac{E_{\rm cl}}{N}=\frac{J}{4}\left(\bm{\lambda}_{\rm A}+\bm{\lambda}_{\rm B}+\bm{\lambda}_{\rm C}-\frac{2\bm{H}}{3J}\right)^2 -\frac{h^2}{9J}-J,\label{eneclassical}
\end{eqnarray}
where $N$ is the number of sites, $\bm{\lambda}_{ \mu}\equiv \langle \Psi^{\rm cl}|\hat{\bm{\lambda}}_{i_\mu}|\Psi^{\rm cl}\rangle$ is an eight-component classical vector of length $\sqrt{4/3}$, and $\bm{H}=(0,0,H,0,0,0,0,0)$. The minimization of $E_{\rm cl}$ is simply achieved when $\overline{\bm{\lambda}_{\mu}}={2\bm{H}}/{9J}$ for $h\leq 3J$. The overline means the average over $\mu={\rm A},{\rm B},{\rm C}$. For $H>3J$, the two conditions $\overline{S^z_{\mu}}=2H/9J$ and $\overline{Q^{z^2}_{\mu}}=0$, cannot be simultaneously satisfied because $Q^{z^2}_{\mu}/\sqrt{3}$ must be larger than $S^{z}_{\mu}-2/3$ from the definition. After some algebra, we found the conditions $\overline{S^z_{\mu}}=H/6J+3/2$ and $\overline{Q^{z^2}_{\mu}}=H/4J-3/4$ in the range of $3J<H<H_{\rm s}$, with $H_{\rm s}=9J$ being the saturation field.

From the above discussion, the magnetization $M\equiv \sum_i \langle \hat{S}^z_i \rangle/N=\overline{S^z_{\mu}}$ is uniquely determined as shown in Fig.~1 (dashed line). However, the specific spin and quadratic orders remain massively degenerate because the number of conditions is smaller than that of variational parameters $d_{i_\mu,\sigma}$.

\subsection{\label{1}Linear flavor-wave excitation spectra}
Here let us supplement the argument on the order-by-disorder selection from the classical degeneracy manifold {on the basis of} linear flavor-wave theory~\cite{Spapanicolaou-88,Slacroix-11,Stoth-12,Sbauer-12}. The linear flavor-wave theory, which is an extension of the spin-wave theory to SU($\mathcal{N}$) systems, gives linear excitation spectra $\omega_\lambda({\bm{k}})$ ($\lambda=1,\cdots, n_{\rm LFW}$) of fluctuations around the mean-field ground state. The number of branches $n_{\rm LFW}$ ($=6$ in the present case) in the reduced Brillouin zone is the product of $\mathcal{N}-1$ and the number of sublattices, and $\bm{k}$ is the quasi-momentum of the {bosonic excitation}  (``flavon'' or ``coloron''). As the calculations are standard and {entirely} similar to those described in several previous papers~\cite{Spapanicolaou-88,Slacroix-11,Stoth-12,Sbauer-12}, we shall present only the results below. 

\begin{figure*}[t]
\includegraphics[scale=0.65]{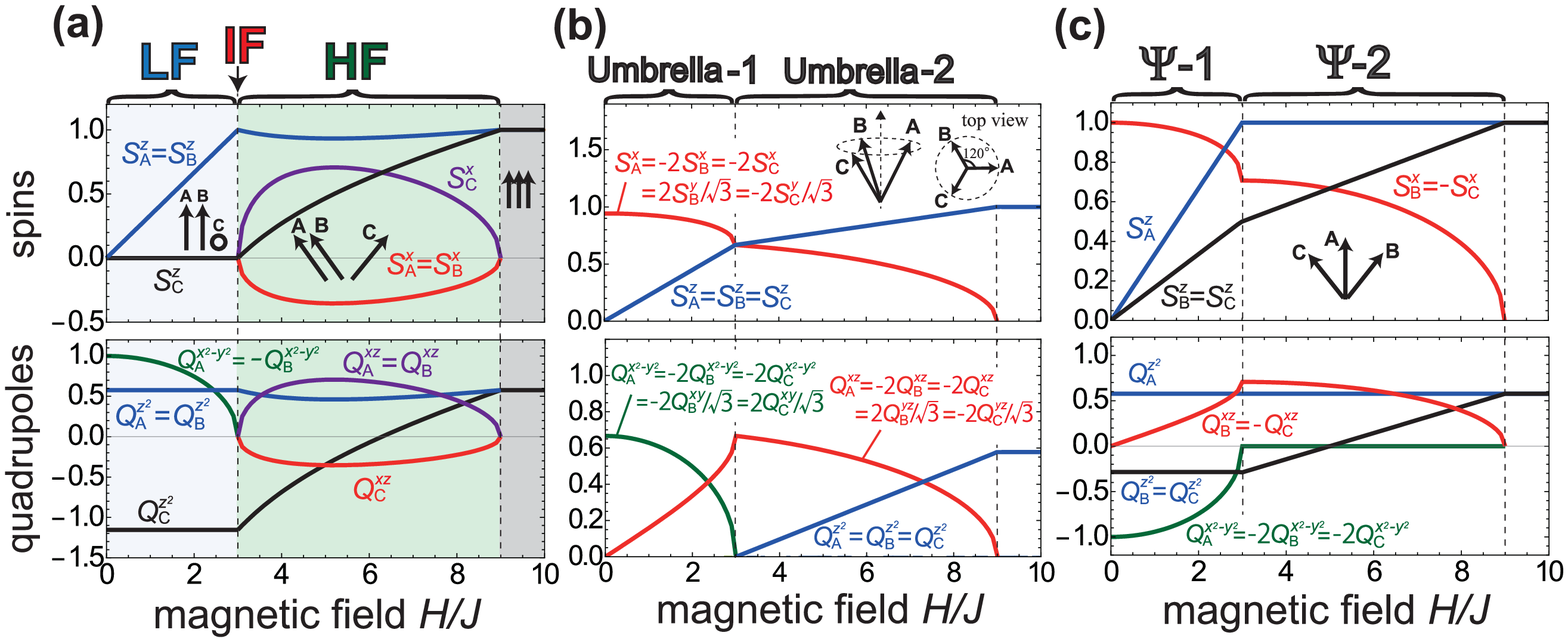}
\caption{\label{figS3}
{Nonzero components of the spin and quadrupolar moments within the mean-field analysis in a fixed gauge with $\langle\hat{Q}_A^{xy} \rangle=\langle\hat{Q}_A^{yz} \rangle=0$ for (a) the sequence of LF, IF, and HF states, (b) the umbrella state, and (c) the $\Psi$ state. The sublattice spin components of each phase are illustrated with arrows. }} 
\end{figure*}
\begin{figure*}[t]
\includegraphics[scale=0.6]{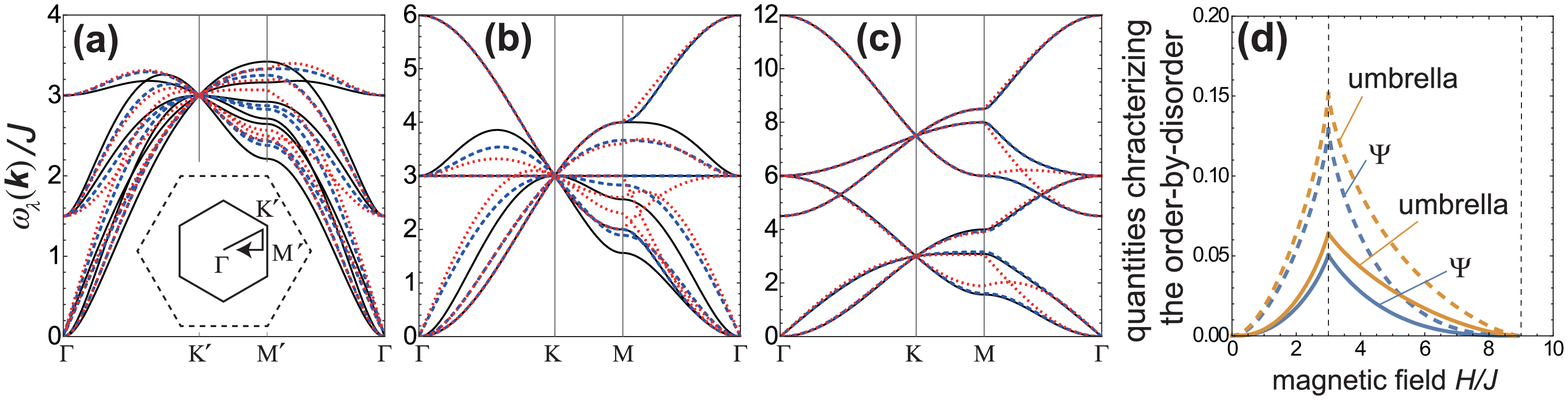}
\caption{\label{figS4}
{Linear flavor-wave excitation spectra around the mean-field solutions of candidate states in the classical ground-state manifold at (a) $H/J=1.5$, (b) $H/J=3$, and (c) $H/J=6$. The black solid curves correspond to the LF state for (a), to IF for (b), and to HF for (c). The red dotted and blue dashed curves are the umbrella and $\Psi$ solutions, respectively. The illustration in (a) shows the reduced Brillouin zone of triangular lattice. (d) The differences in $\frac{1}{2N}\sum_{\bm{k},\lambda} \omega_\lambda({\bm{k}})/J$ (solid curves) and in $\frac{1}{N}\sum_{\bm{k},\lambda} \ln \omega_\lambda({\bm{k}})$ (dashed curves) of the umbrella and $\Psi$ states from those of the sequence of LF, IF, and HF states. }} 
\end{figure*}
For comparison with the states selected by the order-by-disorder mechanism (according to CMF+S in the main text), namely LF, IF, and HF, let us consider the other candidate states (named ``umbrella'' and ``$\Psi$'' following the SU(2) case~\cite{Sstarykh-14,Syamamoto-17}) of highly-symmetric shape. The classical, mean-field values of the spin and quadrupolar components on each sublattice for those candidate states, which are obtained so that Eq.~(\ref{eneclassical}) can be minimized, are shown in Figs.~\ref{figS3}(a-c). The shapes of the umbrella and $\Psi$ states in the spin sector are illustrated in each figure. It is seen that the umbrella and $\Psi$ states {undergo a qualitative change in their quadrupolar sector at the point $H/J=3$ [hence the labels (``Umbrella-1'', ``Umbrella-2'') and (``$\Psi$-1'', ``$\Psi$-2'') in  Figs.~\ref{figS3}(b-c)].}
Figures~\ref{figS4} (a-c) show the comparisons of the linear flavor-wave excitation spectra of them at $H/J=1.5$, 3, and 6. As can be seen, the excitations of the LF, IF, and HF states are softer than the others in each magnetic field range. Quantum fluctuations favor those states with smaller zero-point energy $\frac{1}{2}\sum_{\bm{k},\lambda} \omega_\lambda({\bm{k}})$, while thermal fluctuations tend to minimize $\sum_{\bm{k},\lambda} \ln \omega_\lambda({\bm{k}})$~\cite{Smila-15} to select the one that has the largest entropy. Within the linear flavor-wave theory, {both types of} fluctuations choose the sequence of LF, IF, and HF states [Fig~\ref{figS4}(d)], which is confirmed by the CMF+S and semi-classical Monte-Carlo analyses in the main text.

\newpage
\subsection{\label{2}Technical details of the semi-classical {Monte-Carlo} analysis}
In the main text, we employ the semi-classical Monte-Carlo simulations~\cite{Sstoudenmire-09} on $L\times L$ rhombic clusters under periodic boundary conditions
, since the fully-quantum Monte-Carlo method suffers from the so-called sign problem for frustrated quantum systems. First, we assume that the wave function of the entire system is described as a direct products of local wave functions {as in the MF approximation [Eq.~(\ref{direct_product})], although the three-sublattice ansatz is not assumed. }
The total energy of the system is given by
\begin{eqnarray}
E_{\rm cl}(\{ \bm{d}_i\})=\langle \Psi^{\rm cl}|\hat{\mathcal{H}}|\Psi^{\rm cl}\rangle=J\sum_{\langle i,j\rangle}\left(|\bm{d}_i^\ast\cdot\bm{d}_j|^2-\frac{1}{3}\right)-H\sum_i\left(|d_{i,1}|^2-|d_{i,-1}|^2\right)\label{sclene}
\end{eqnarray}
within the direct-product approximation. We first set the initial values of $\bm{d}_{i}$ on the entire lattice sites to $L\times L$ complex random vectors distributed homogeneously on the sphere of radius one in 3 (real) +3 (imaginary) dimensions. Starting with the initial state, we perform the standard Metropolis local updates of $\bm{d}_{i}$ to generate a sequence of states weighted by the probability proportional to the Boltzmann factor $\exp (- E_{\rm cl}(\{ \bm{d}_i\})/k_{\rm B} T)$. Typical simulations contain $10^5$ and $2\times 10^6$ Monte-Carlo steps for the thermalization of the state and the samplings of physical quantities, respectively. One Monte-Carlo step consists of one Metropolis sweep over all sites followed by two ``relaxation acceleration'' sweeps (which will be explained in Sec.~{D}). 

The quantum-mechanical expectation values of the local spin and quadrupolar moments can be calculated by
\begin{eqnarray}
\lambda^A_i\equiv \langle \psi_i|\hat{\lambda}^A_i |\psi_i\rangle = \sum_{\sigma,\sigma^\prime} \langle \sigma_i|\hat{\lambda}^A_i |\sigma^\prime_i\rangle d^\ast_{i,\sigma}d_{i,\sigma^\prime}
\end{eqnarray}
for a given site with vector $\bm{d}_i$. The eight components of the vector $\hat{\bm{\lambda}}_i$ correspond to the spin components ($\hat{S}_i^x,\hat{S}_i^y,\hat{S}_i^z$) for $A=1,2,3$ and quadrupolar components ($\hat{Q}^{x^2-y^2}_i,\hat{Q}^{z^2}_i,\hat{Q}^{xy}_i,\hat{Q}^{yz}_i,\hat{Q}^{xz}$) for $A=4,5,\cdots,8$, respectively, as in the main text. To discuss the spontaneous symmetry breaking, we calculate the correlation lengths of the diagonal and  transverse components:
\begin{eqnarray}
\xi^{\parallel}_{(1,2)}=\frac{\sqrt{3}L}{4\pi}\sqrt{\frac{\mathcal{S}^{\parallel}_{(1,2)}({\bm{Q}_{\rm K}})}{\mathcal{S}^{\parallel}_{(1,2)}({\bm{Q}_{\rm K}}+(0,4\pi/\sqrt{3}L))}-1}~~{\rm and}~~\xi^{\perp}_{(1,2)}=\frac{\sqrt{3}L}{4\pi}\sqrt{\frac{\mathcal{S}^{\perp}_{(1,2)}({\bm{Q}_{\rm K}})}{\mathcal{S}^{\perp}_{(1,2)}({\bm{Q}_{\rm K}}+(0,4\pi/\sqrt{3}L))}-1}
\end{eqnarray}
with the structure factors 
\begin{eqnarray}
\mathcal{S}^{\parallel}_{(1)}(\bm{k})&=&\frac{1}{L^2}\sum_{i,j}\langle\!\langle S_i^z S_j^z\rangle\!\rangle_T e^{-i\bm{k}\cdot(\bm{r}_i-\bm{r}_j)},~~\mathcal{S}^{\parallel}_{(2)}(\bm{k})=\frac{1}{L^2}\sum_{i,j}\langle\!\langle Q_i^{z^2} Q_j^{z^2}\rangle\!\rangle_T e^{-i\bm{k}\cdot(\bm{r}_i-\bm{r}_j)},\nonumber\\
\mathcal{S}^{\perp}_{(1)}(\bm{k})&=&\frac{1}{L^2}\sum_{i,j}\frac{\langle\!\langle Q_i^{x^2-y^2} Q_j^{x^2-y^2}+ Q_i^{xy} Q_j^{xy}\rangle\!\rangle_T}{2} e^{-i\bm{k}\cdot(\bm{r}_i-\bm{r}_j)},~~{\rm and}\nonumber\\
\mathcal{S}^{\perp}_{(2)}(\bm{k})&=&\frac{1}{L^2}\sum_{i,j}\frac{\langle\!\langle S_i^{x} S_j^{x}+ S_i^{y} S_j^{y}+ Q_i^{yz} Q_j^{yz}+ Q_i^{xz} Q_j^{xz}\rangle\!\rangle_T}{2} e^{-i\bm{k}\cdot(\bm{r}_i-\bm{r}_j)}.
\label{st}
\end{eqnarray}
Here, $\langle\!\langle\cdots\rangle\!\rangle_T$ means the thermal average in terms of Monte-Carlo samplings and the ordering vector ${\bm{k}}=\bm{Q}_{\bf K}\equiv(4\pi/3,0)$ corresponds to the three-sublattice order shown as the inset of Fig. 2(a).

The stiffness $\rho_{S^z}(T)$ for a twist generated by the unitary transformation, $\hat{U}_{S^z}(q)\equiv \exp[i q \sum_i x_i \hat{S}^z_i]$, is defined in the standard way as the second derivative of the free energy per unit area with respect to the twist angle $q$: 
\begin{eqnarray}
\rho_{S^z}(T)&=&\frac{1}{L^2\Delta S}\left\langle\!\!\!\left\langle\frac{\partial^2\langle \hat{U}_{S^z}(q)\hat{\mathcal H}\hat{U}_{S^z}^\dagger(q)\rangle}{\partial q^2}\Bigg|_{q=0}\right\rangle\!\!\!\right\rangle_T-\frac{1}{k_{\rm B}T}\left\langle\!\!\!\left\langle\Bigg(\frac{\partial\langle \hat{U}_{S^z}(q)\hat{\mathcal H}\hat{U}_{S^z}^\dagger(q)\rangle}{\partial q}\Bigg|_{q=0}\Bigg)^2\right\rangle\!\!\!\right\rangle_T\nonumber\\
&=&-\frac{2}{\sqrt{3}L^2}\left\langle\!\!\!\left\langle\frac{J}{2}\sum_{\langle i,j\rangle}(x_i-x_j)^2\left(S_i^xS_j^x+S_i^yS_j^y+4(Q^{x^2-y^2}_iQ^{x^2-y^2}_j+Q^{xy}_iQ^{xy}_j)+Q^{yz}_iQ^{yz}_j+Q^{xz}_iQ^{xz}_j\right)\right\rangle\!\!\!\right\rangle_T\nonumber\\
&&-\frac{2}{\sqrt{3}L^2k_{\rm B}T}\left\langle\!\!\!\left\langle\left[\frac{J}{2}\sum_{\langle i,j\rangle}(x_i-x_j)\left(S_i^xS_j^y-S_i^yS_j^x+2(Q^{x^2-y^2}_iQ^{xy}_j-Q^{xy}_iQ^{x^2-y^2}_j)\right.\right.\right.\right.\nonumber\\
&&~~~~~~~~~~~~~~~~~~~~~~~~~~~~~~~~~~~~~~~~~~~~~~~~~~~~~~~~~~~~~~~~~~-(Q^{yz}_iQ^{xz}_j-Q^{xz}_iQ^{yz}_j)\Big)\Bigg]^2\Bigg\rangle\!\!\!\Bigg\rangle_T,
\end{eqnarray}
where $\Delta S=\sqrt{3}/2$ is the area per site. Here, we choose the twist direction to be parallel to the $x$-axis, although the value of $\rho_{S^z}(T)$ does not depend on this choice for $L\rightarrow \infty$. In a similar way, the stiffness $\rho_{P^z_+}(T)$ regarding $\hat{U}_{P^z_{+}}(q)\equiv \exp[i q \sum_i x_i \hat{P}^z_{+,i}]$ with $\hat{P}^z_{+,i}\equiv \frac{1}{2}\hat{S}_i^z+\frac{\sqrt{3}}{2}\hat{Q}_i^{z^{2}}$ is defined as
\begin{eqnarray}
\rho_{P^z_+}(T)&=&-\frac{2}{\sqrt{3}L^2}\left\langle\!\!\!\left\langle\frac{J}{2}\sum_{\langle i,j\rangle}(x_i-x_j)^2\left(4(P_{+,i}^{x}P_{+,j}^{x}+P_{+,i}^yP_{+,j}^y)+P_{-,i}^{x}P_{-,j}^{x}+P_{-,i}^yP_{-,j}^y+Q^{x^2-y^2}_iQ^{x^2-y^2}_j+Q^{xy}_iQ^{xy}_j\right)\right\rangle\!\!\!\right\rangle_T\nonumber\\
&&-\frac{2}{\sqrt{3}L^2k_{\rm B}T}\left\langle\!\!\!\left\langle\left[\frac{J}{2}\sum_{\langle i,j\rangle}(x_i-x_j)\left(2(P_{+,i}^{x}P_{+,j}^{y}-P_{+,i}^yP_{+,j}^x)-(P_{-,i}^{x}P_{-,j}^{y}-P_{-,i}^yP_{-,j}^x)\right.\right.\right.\right.\nonumber\\
&&~~~~~~~~~~~~~~~~~~~~~~~~~~~~~~~~~~~~~~~~~~~~~~~~~~~~~~~~~~~~~~~~~~+Q^{x^2-y^2}_iQ^{xy}_j-Q^{xy}_iQ^{x^2-y^2}_j\Big)\Bigg]^2\Bigg\rangle\!\!\!\Bigg\rangle_T,
\end{eqnarray}
where $\hat{P}^x_\pm\equiv (\hat{S}^x\pm\hat{Q}^{xz})/\sqrt{2}$ and $\hat{P}^y_\pm \equiv (\hat{S}^y\pm\hat{Q}^{yz})/\sqrt{2}$.

\subsection{\label{3}Numetical data of the {semi-classical Monte-Carlo} simulations}
\begin{figure}[t]
\includegraphics[scale=0.58]{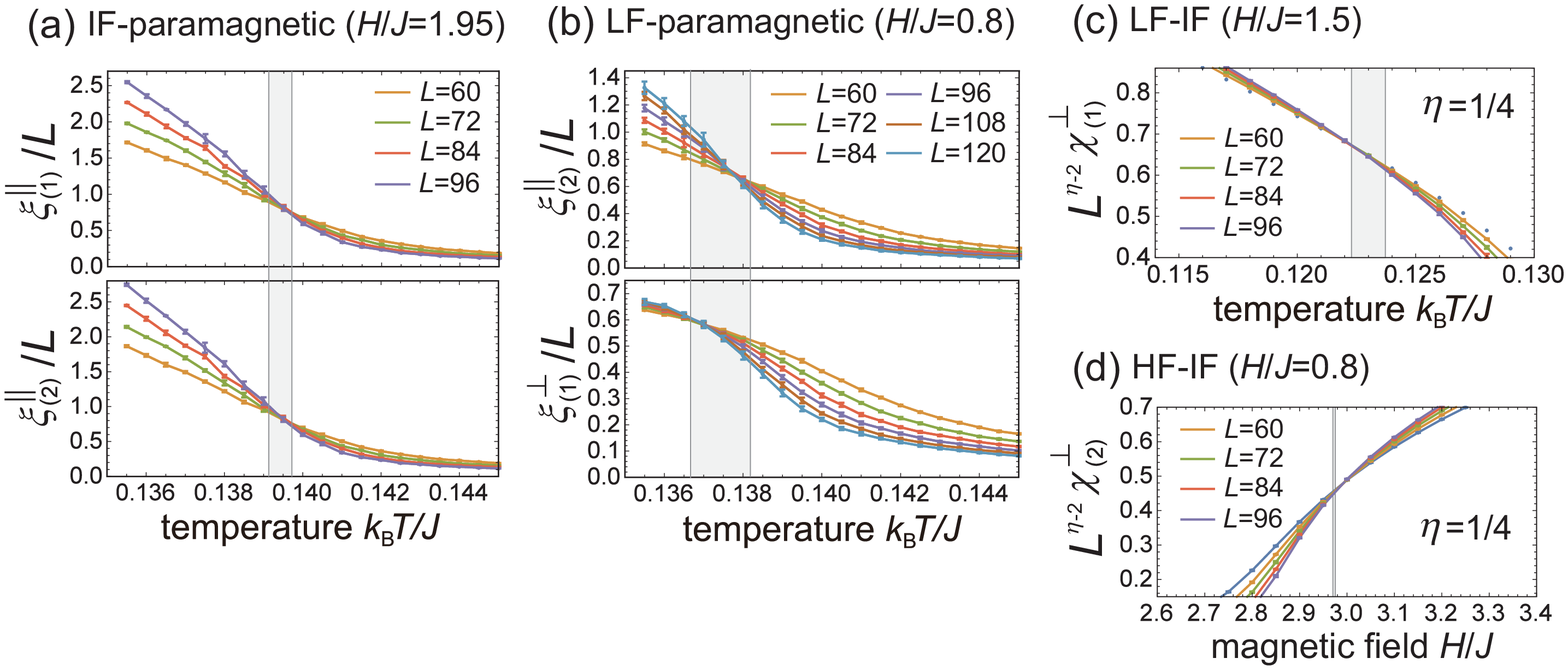}
\caption{\label{figS1}
Typical examples of the numerical data for the scaling analyses that determine the finite-temperature phase diagram shown in Fig. 3 of the main text. The error bars assigned to  each data point are estimated as the square root of variance of about 10 independent {semi-classical Monte-Carlo} simulations. The shaded bands indicate the estimated transition points with their error bar. } 
\end{figure}
Here we present some numerical data for the standard scaling analyses performed to determine the phase boundaries of Fig. 3 of the main text. All the three (LF, IF, and HF) phases possess a three-sublattice order in the diagonal components, $S^z$ and $Q^{z^2}$. Therefore, the transition points to the paramagnetic phase can be identified by the divergence of $\xi^{\parallel}_{(1,2)}$. Figure~\ref{figS1} (a) shows a typical example of the transition from the IF to paramagnetic phase. The curves of the scaled correlation length $\xi^{\parallel}_{(1,2)}/L$ for different linear sizes $L$ cross each other at a critical point, {within the error bar estimated from the square root of the variance over about 10 {semi-classical Monte-Carlo} simulations.} We plot in Fig.3 the crossing points in $\xi^{\parallel}_{(2)}/L$, which are slightly less size-dependent, as the phase boundary from the ordered to the paramagnetic states.

Note that, in the standard Berezinskii-Kosterlitz-Thouless transition of the 2D XY model with rotational symmetry, the scaled correlation length of the transverse (XY) components does not exhibit an isolated critical (crossing) point but a finite critical range from $T=0$ with a constant value independent of $L$. Interestingly, the scaled correlation length of the transverse components, $\xi^{\perp}_{(1)}/L$ ($\xi^{\perp}_{(2)}/L$) in the present case shows a crossing behavior in the vicinity of the transitions from the LF (HF) to parmagnetic transitions, in spite of the continuous nature of the rotational symmetries of the system around the $S^z$ and $Q^{z^2}$ axes [see an example for the LF-paramagnetic transition in Fig.~\ref{figS1}(b)]. This may be attributed to the combined effect of the simultaneous discrete (diagonal) and continuous (transverse) symmetry breakings. 
A similar (apparent) crossing behavior of the scaled correlation length for the transverese components has been reported in previous studies on some related 2D models with combined discrete and continuous symmetry breakings~\cite{Sseabra-11}. The crossing points in the scaled correlation lengths of the diagonal and transverse components are located at almost the same position (within the error bar) as seen in Fig. 3 of the main text.

From the LF to the IF (HF to IF) phase, the topological transition associated with the unbinding of pairs of half-vortex and half-antivortex in the plane of $Q^{x^2-y^2}$ and $Q^{xy}$ ($P_+^{x}$ and $P_+^{y}$) occurs, as explained in the main text. In this case, the corresponding scaled correlation length does not exibit an isolated critical point. Therefore, to locate the topological transition points, we perform the scaling analysis on the susceptibilities of the corresponding quantities:
\begin{eqnarray}
\chi^{\perp}_{(1)}=\frac{J}{k_{\rm B}T}\mathcal{S}^{\perp}_{(1)}(\bm{Q}_{\rm K})~~~{\rm and}~~~\chi^{\perp}_{(2)}=\frac{J}{k_{\rm B}T}\mathcal{S}^{\perp}_{(2)}(\bm{Q}_{\rm K}),
\end{eqnarray}
which obey the following scaling relations:
\begin{eqnarray}
\chi^{\perp}_{(1,2)}=L^{2-\eta} \tilde{\chi}^{\perp}_{(1,2)} (t L^{1/\nu})\label{scale}
\end{eqnarray}
with unknown universal functions $\tilde{\chi}^{\perp}_{(1,2)}$ of $t=(T-T_{{\rm c}})/T_{{\rm c}}$. Here, $\eta$ and $\nu$ are the correlation function and  correlation length critical exponents, respectively. At the LF-to-IF (HF-to-IF) topological phase transition, $\chi^{\perp}_{(1)}$ ($\chi^{\perp}_{(2)}$) is expected to  scale with the exact {Berezinskii-}Kosterlitz-Thouless exponent $\eta= 1/4$~\cite{Sgvozdikova-11}. According to Eq.~(\ref{scale}), the quantities $L^{\eta-2} \chi^{\perp}_{(1,2)}$ become size-independent at the corresponding transition points $t=0$ with $\eta= 1/4$. Figure~\ref{figS1}(c) and~\ref{figS1}(d) show typical examples of the scaling analysis performed to determine the LF-IF and HF-IF boundaries, respectively, plotted in Fig. 3.

\subsection{\label{4}Relaxation acceleration techniques}
Technical details of the ``relaxation acceleration'' techniques we introduced in the main text are presented below. The explicit form of the local effective Hamiltonian on site $i$ within the direct-product approximation is given by 
\begin{eqnarray}
\hat{\mathcal{H}}^{\rm loc}_i\equiv (\otimes_{j\neq i}\langle \psi_j|)\hat{\mathcal{H}}(\otimes_{j\neq i}|\psi_j\rangle)=\frac{J}{2}\sum_{j\in {\rm NN}_i}\left(\bm{S}_j\cdot\hat{\bm{S}}_i+\bm{Q}_j\cdot\hat{\bm{Q}}_i\right)-H\hat{S}_i^z,
\end{eqnarray}
where the sum $\sum_{j\in {\rm NN}_i}$ runs over all nearest-neighbor sites of site $i$ and the terms independent on the local state on site $i$ are ignored. Note that the energy of the system within the direct-product approximation [Eq.~(\ref{sclene})] is preserved under the local unitary transformation $e^{i c \hat{\mathcal{H}}^{\rm loc}_i}|\psi_i\rangle$ with $c$ being a real number. Using the local unitary transformations combined with the Metropolis updates, one can avoid the problem of trapping into local minima and significantly improve the decorrelation between the adjacent Monte-Carlo samples in the Markov chain. The operator $e^{i c \hat{\mathcal{H}}^{\rm loc}_i}$ can be described as the matrix exponential of a $3\times 3$ matrix on the local state basis (\ref{direct_product}). The numerical cost for computing the matrix exponential can be reduced by using the spectral decomposition and the analytical form of the eigenvalues of the $3\times 3$ Hermitian matrix $\hat{\mathcal{H}}^{\rm loc}_i$~\cite{Scopp-06}.

\begin{figure}[t]
\includegraphics[scale=0.6]{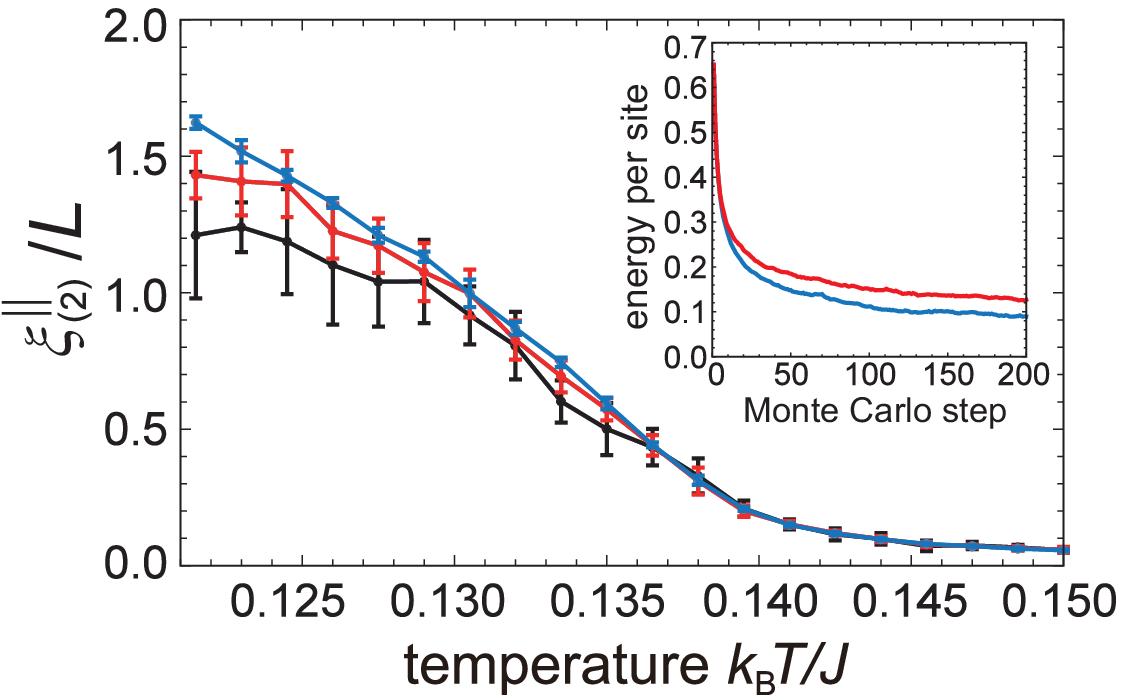}
\caption{\label{figS2}
Comparison of the results with and without {relaxation-acceleration}. The values of the scaled correlation length $\xi^{\parallel}_{(2)}/L$ for $H/J=0.5$ and $L=96$ obtained by the {semi-classical Monte-Carlo} simulations (i) for $10^6$ samples without {relaxation-acceleration} (black), (ii) for $3\times 10^6$ samples without {relaxation-acceleration} (red), and (iii) for $10^6$ samples with {relaxation-acceleration} (blue) are plotted. The error bars are estimated from the square root of the variance over nine independent {semi-classical Monte-Carlo} simulations. The inset shows the energy of the system per site for the first 200 Monte-Carlo steps in the thermalization processes with (blue) and without (red) {relaxation-acceleration}.} 
\end{figure}
In Fig.~\ref{figS2}, we compare the calculated values of the scaled correlation length $\xi^{\parallel}_{(2)}/L$ for $H/J=0.5$ and $L=96$, as am example, obtained by the {semi-classical Monte-Carlo} simulations (i) for $10^6$ samples without {relaxation-acceleration} (black), (ii)  for $3\times 10^6$ samples without {relaxation-acceleration} (red), and (iii) for $10^6$ samples with {relaxation-acceleration} (blue). Here, one Monte-Carlo step consists of two sweeps of the {relaxation-acceleration} operations over all sites following one Metropolis update sweep, and the sampling of the physical quantities for calculating the thermal average is performed at every Monte-Carlo step. The arbitrary real number $c$ is chosen to be uniformly distributed random numbers in $[-\pi f_{\rm n}^{-1},\pi f_{\rm n}^{-1}]$ with $f_{\rm n}$ being the Frobenius norm of the matrix form of $\hat{\mathcal{H}}^{\rm loc}_i$. As can be seen in Fig.~\ref{figS2}, the error bars of the data are clearly diminished owing to the {relaxation-acceleration} operations, even in comparison of (ii) and (iii) with a three times difference in those sample numbers, which take roughly same computation time. This indicates the reduction of the autocorrelation between the samples. The inset shows the {semi-classical Monte-Carlo} thermalization processes from an initial state with uniformly distributed random vectors $\bm{d}_i$ on the entire lattice sites. It can be seen that the case with {relaxation-acceleration} shows faster convergence to the thermal equilibrium.

The acceleration and decorrelation of the {Monte-Carlo} updates by the {relaxation-acceleration} technique are expected to become increasingly important for models with  higher symmetry, such as SU($\mathcal{N}\ge 3$).

\end{document}